\documentclass[prd,aps,nofootinbib,onecolumn,showpacs,10pt]{revtex4-1}
\usepackage{graphicx,epsfig}

\usepackage{epsf}
\usepackage{graphicx}
\def\ds{\displaystyle}

\begin{document}
\def\ds{\displaystyle}
\def\beq{\begin{equation}}
\def\eeq{\end{equation}}
\def\bea{\begin{eqnarray}}
\def\eea{\end{eqnarray}}
\def\beeq{\begin{eqnarray}}
\def\eeeq{\end{eqnarray}}
\def\ve{\vert}
\def\vel{\left|}
\def\ver{\right|}
\def\nnb{\nonumber}
\def\ga{\left(}
\def\dr{\right)}
\def\aga{\left\{}
\def\adr{\right\}}
\def\lla{\left<}
\def\rra{\right>}
\def\rar{\rightarrow}
\def\nnb{\nonumber}
\def\la{\langle}
\def\ra{\rangle}
\def\bdll{$B_c \rightarrow D_{q'}^*l^+ l^-$}
\def \xjga{$\chi_{c0} \rar J/\psi \gamma $~}
\def \xUga{$\chi_{b0} \rar \Upsilon \gamma $~}
\def\ba{\begin{array}}
\def\ea{\end{array}}
\title{  \bf   Study of $\chi_{c0}(1P) \rar J/\psi \gamma $ and $\chi_{b0}(1P) \rar \Upsilon(1S) \gamma $ decays via QCD sum rules}
\author{V. Bashiry $^{*1}$ \\
  $^{*}$Cyprus International University, Faculty of Engineering, Nicosia, Northern Cyprus,\\ Mersin 10, Turkey\\
$^1$e-mail:bashiry@ciu.edu.tr}

\begin{abstract}
In this study we present the first theoretical calculation of the form factor of the $\chi_{c0}(1P) \rar J/\psi \gamma $ and $\chi_{b0}(1P) \rar \Upsilon(1S) \gamma $ decays in the frame work of QCD sum rules. We also find  branching ratio   ${\cal B}_r(\chi_{c0}(1P) \rar J/\psi \gamma)=(1.07\times \pm 0.14) \times 10^{(-2)} $ which is in agreement with the experimental data. Furthermore, we estimate the $\Gamma_{tot}(\chi_{b0}(1P)) =5.5\pm 0.5 $~MeV, where experimental bound for full width of $\chi_{b0}$ is $\Gamma_{tot}(\chi_{b0}(1P)) <6 $~MeV.
\end{abstract}

\maketitle

\section{Introduction}
Heavy quarkonium states( like  $b\bar{b}$ and $c\bar{c}$) and their decay modes offer a laboratory to
study the strong interaction in the non-perturbative regime. Charmonium in particular has
served as a calibration tool for the corresponding techniques and models \cite{1}.
Heavy quarkonium states can have many bound states and decay channels used to study and determine  different parameters of Standard
Model(SM) and QCD from the theoretical perspective. In particular, the calculation  of bottomonium masses\cite{Veliev:2011kq}, total widths, coupling constants\cite{Veliev:2010vd,Sundu:2011vz, Wang:2012kw,Rodrigues:2010ed} and branching ratio can serve as  benchmarks for the low energy predictions of QCD.
In addition, the theoretical calculations on the branching ratio of radiative decays of heavy quarkonium states, are relatively clean with respect to the hadronic or semileptonic decays,  and their comparison with experimental data could provide inportant insights into their nature and hyperfine interaction.  In this regard,  we present the first theoretical study on the branching ratio of inclusive  \xjga and \xUga decays. Note that in order to calculate the branching ratio we have to get information about the masses and decay
constants of the participating particles. It is worth  mentioning that the masses can be obtained either  by means of the  experimental
results i.e,  the Particle Data Group\cite{PDG2012} or by the theoretical methods . The decay constants, on
the other hand, can be calculated theoretically via different non-perturbative methods. In this respect, masses and decay constants and spectrums of heavy qurakonium states are calculated in the various approaches( see for instance \cite{pdg1}-\cite{Bashiry:2011pp}). Here, firstly, we calculate the form factor of \xjga and \xUga decays in the framework of three-ponit QCD  sum rules, secondly we calculate the branching ratio of inclusive  \xjga and \xUga decays.
In section 2, we introduce the QCD sum rules technique  for the form factors of inclusive  \xjga and \xUga decays.  Last section is devoted to  the
numerical analysis and discussion.

\section{QCD Sum Rules for the form factors}

The three-point correlation function associated with the $\chi_{c0}(1P) \rar J/\psi \gamma $ and $\chi_{b0}(1P) \rar \Upsilon(1S) \gamma $ vertex is given by
\begin{eqnarray}\label{CorrelationFunc1}
\Pi_{\mu}^=i^2 \int d^4x~d^4y~
e^{-ip\cdot x+ip^{\prime}\cdot y}~ {\langle}0| {\cal T}\left(j^{V}_{\mu}(y)j^{em}_\nu(x)\bar{j}^{S}(0)~\right)|0{\rangle}
\end{eqnarray}

%
%
 where ${\cal T}$ is  the time ordering operator and $q$ is momentum of photon. Each meson  and photon field can be described in terms of
the quark field operators as follows:
\begin{eqnarray}\label{mesonintfield}
j^{V}_{\mu}(y)&=& \overline{c(b)}(y)\gamma_{\mu}c(b)(y)
\nonumber \\
j^{S}(x)&=& \overline{c(b)}(x)c(b)(x)\nonumber \\
j^{em}_{\mu}(x)&=&Q_{c(b)} \overline{c(b)}(x)\gamma_{\mu}c(b)(x)
\end{eqnarray}

We calculate the correlation function Eq. (\ref{CorrelationFunc1}) in two different methods. In phenomenological or physical side, it can be evaluated  in terms of hadronic parameters such as masses, decay constants and form factors. In theoretical or  QCD  side, on the other hand,
it is calculated in terms of QCD parameters, which are quark and gluon degrees of freedom, by the help
of the operator product expansion (OPE) in deep Euclidean region. Equating the structure calculated in two different approaches of the same correlation function, we get  a relation between hadronic parameters and QCD degrees of freedom. Finally, we apply double Borel transformation with respect to the momentum of initial and final mesons($p^2$ and $p'^2$). This final operation suppresses the  contribution of the higher states and continuum.
\subsection{Phenomenological side}
We insert the complete sets of appropriate vector meson($|V\rangle\langle V|$) and scalar meson($|S\rangle\langle S|$) states (regarding the conservation of  the quantum numbers of corresponding interpolating currents) inside correlation functions Eq.(\ref{CorrelationFunc1}). Here, vector state is either $j/\psi$ or $\Upsilon$ and scalar state is $\chi_{c(b)}$ state.   After integrating   over the $x$ and $y$,
we get the following result for the correlation function Eq.(\ref{CorrelationFunc1}):
\begin{eqnarray}\label{CorrelationFuncPhys1}
\Pi_{\mu}&=&\frac{{\langle}0|j^{S}(x)|S{\rangle} {\langle}S|j^{em}_{\nu}(x)|V{\rangle} {\langle}V|\bar{j}^V_{\mu}|0{\rangle}}
{(m_{V}^2-p^{\prime^2})(m_{S}^2-p^2)} +...
\end{eqnarray}
where .... contains the contribution of the higher and continuum states with the same quantum numbers .

The  matrix elements of  the  above
equation are  related to the  hadronic parameters  as follows:
\begin{eqnarray}\label{transitionamp}
{\langle}0|j^{V}_{\mu}(x)|V{\rangle}&=&m_{V}f_{V}\epsilon^\prime_{\mu}
\nonumber\\
{\langle}S|\bar{j}^S|0{\rangle}&=& i m_{S}f^\ast_S
\nonumber\\
{\langle}S|j^{em}_{\nu}(x)|V{\rangle}&=&e F(q^2=0)\{(p^\prime \cdot q)\epsilon^{\prime}_{\nu}  -(q\cdot \epsilon^{\prime})p^\prime_{\nu}\}
\end{eqnarray}
where $F(q^2)$ is the form factor of transition  and $\epsilon^{\prime}$ is the polarization vector associated with the vector meson. Using
Eq. (\ref{transitionamp}) in Eq. (\ref{CorrelationFuncPhys1}) and considering the summation over polarization vectors via,
\begin{eqnarray}\label{polvec}
\epsilon_\nu\epsilon^*_\theta=-g_{\nu\theta},\nonumber\\
\epsilon'_j\epsilon^{'*}_\mu=-g_{j\mu}+\frac{p^\prime_{j}p^\prime_{\mu}}{m_{V}^2},
\end{eqnarray}
the  result of the phenomenological or physical side is as follows:
\begin{eqnarray}\label{CorrelationFuncPhys2}
\Pi_{\mu}&=&-\frac{e m_{V}f_{V}m_{S}f^\ast_S}{(m_{V}^2-p^{\prime^2})(m_{S}^2-p^2)} F(0)(p^\prime \cdot q)g_{\mu\nu}+...
\end{eqnarray}
 We  are going to  compare the coefficient of   $g_{\mu\nu}$ structure  for further calculation  from different approaches of the correlation functions.
\subsection{Theoretical(QCD) side}
Theoretical side consists of perturbative(bare loop see fig. (\ref{bare}) and non-perturbative
 parts(the contributions of two gluon condensate diagrams fig. (\ref{gluon})  ). We calculate it in the deep Euclidean space ($p^2\rightarrow-\infty$ and ${p^{\prime}}^2\rightarrow-\infty$).
We consider this side as:
\begin{eqnarray}\label{CorrelationFuncQCD}
\Pi_{\mu}(p^{\prime},p)&=&
\left(\Pi_{per}+\Pi_{nonper}\right)(p^\prime \cdot q)g_{\mu\nu} ,
\end{eqnarray}
\subsubsection{Bare loop}
The perturbative part is a double dispersion
integral as follows:
\begin{figure}[h!]
\begin{center}
\includegraphics[width=12cm]{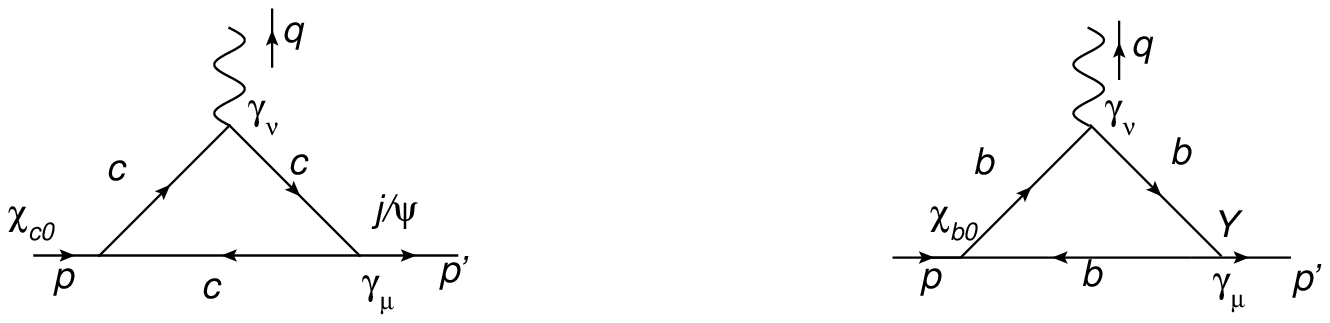}
\end{center}
\caption{The bareloop diagram  for the \xjga
and \xUga decays; } \label{bare}
\end{figure}
\begin{eqnarray}\label{CorrelationFuncQCDPert}
\Pi_{per}&=&-\frac{1}{4 \pi^{2}} \int ds^{\prime} \int ds
\frac{\rho(s,s^{\prime}, q^2)}{(s-p^2)
(s^{\prime}-{p^{\prime}}^2)}+\mbox{subtraction terms},
\end{eqnarray}

where, $\rho(s,s^{\prime}, q^2) $ is  called spectral density.
We aim to  evaluate the spectral density with the help of  the bare
loop diagram in Fig.(\ref{bare}). One of the generic methods to  calculate this
bare loop integral is  the Cutkosky method, where  the quark propagators  of Feynman integrals are replaced by the Dirac
delta functions:
\begin{equation}\label{*}
\frac{1}{q^2-m^2}\rightarrow (- 2\pi i) \delta(q^2-m^2).
\end{equation}
Then, using the Cutkosky method we get spectral density as:
\begin{eqnarray}\label{SpecDenstD}
\rho(s,s^{\prime},q^2)&=&\frac{2m_{c(b)} N_c(-4m_{c(b)}^2+q^2+s-s^\prime)}{3\lambda^{1/2}(s,s^{\prime},q^2)(q^2+s-s^\prime)},
\end{eqnarray}
 where $\lambda(a,b,c)=a^2+b^2+c^2-2ac-2bc-2ab$ and $N_c=3$ is the color number.
 Note that, since three $\delta$ functions of integrand must vanish simultaneously, the physical regions in the $s$ - $s^{\prime}$ plane must satisfy  the following inequality:
\begin{eqnarray}\label{fsspBoffshell}
-1\leq
f(s,s^{\prime})=\frac{s ( q^2 + s - s^{\prime})}{\lambda^{1/2}(m_{c(b)}^2,m_{c(b)}^2,s)
\lambda^{1/2}(s,s^{\prime},q^2)}\leq 1,
\end{eqnarray}

\subsubsection{Two Gluon Condensates}
\begin{figure}[h!]
\begin{center}
\includegraphics[width=12cm]{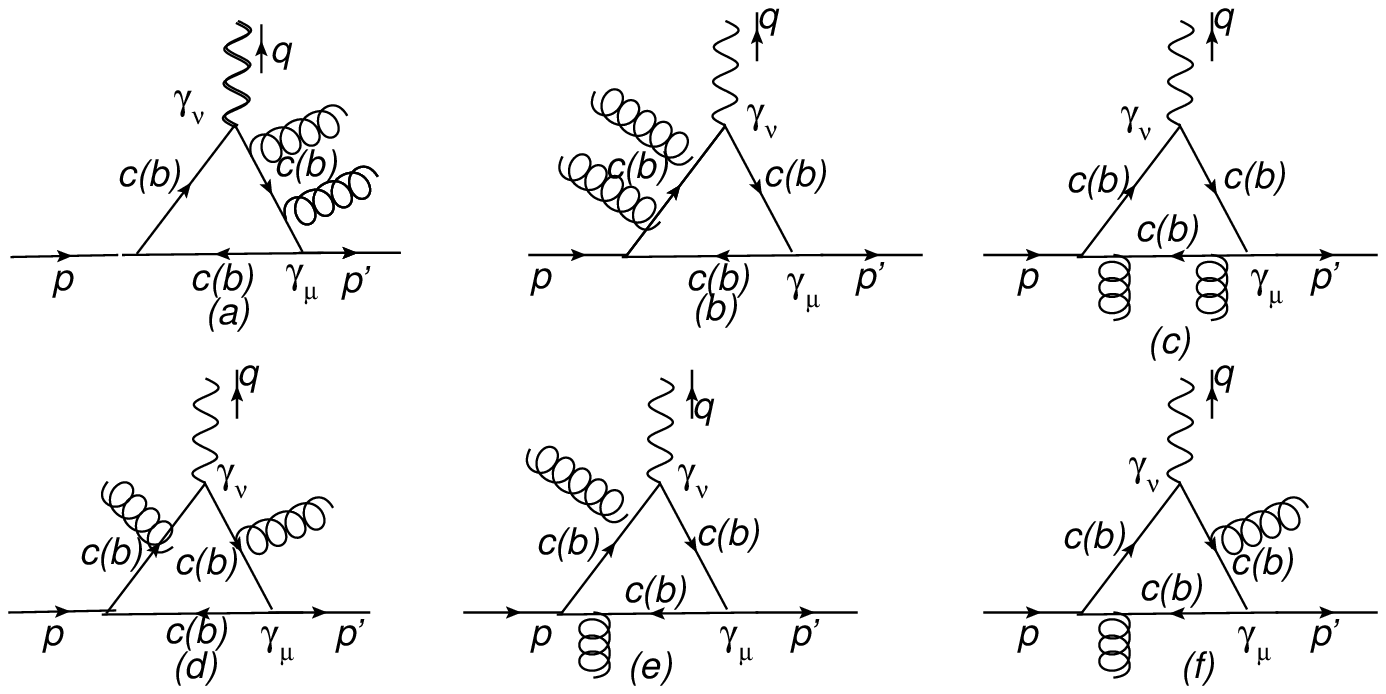}
\end{center}
\caption{Two gluon condensate diagram as a radiative corrections  for the \xjga
and \xUga decays; } \label{gluon}
\end{figure}

 We consider  the two gluon condensate diagrams. Note that, we do not add the heavy quarks condensates, because  the heavy quark
contributions are suppressed by the inverse of the heavy quark mass, so they can be safely neglected. Now, as a nonperturbative part,
 we must add contributions coming  from the gluon condensates presented in (a), (b), (c), (d), (e) and (f) in
Fig. (\ref{gluon}).
These diagrams are calculated  in the Fock--Schwinger fixed--point gauge\cite{R7320,R7321,R7322}  where, the  vacuum gluon field is  as follows:
\begin{eqnarray}\label{Amu}
A^{a}_{\mu}(k')=-\frac{i}{2}(2 \pi)^4 G^{a}_{\rho\mu}(0)\frac{\partial} {\partial k'_{\rho}}\delta^{(4)}(k'),
\end{eqnarray}
where $k'$ is the gluon momentum and  $A_\mu^a$ is the gluon field. In addition, the  quark-gluon-quark vertex is used as:
\begin{eqnarray}\label{qgqver}
\Gamma_{ij\mu}^a=ig\gamma_\mu
\left(\frac{\lambda^{a}}{2}\right)_{ij},
\end{eqnarray}

We come across the following integrals in calculating the gluon condensate contributions
\cite{Aliev3, R7323}: \bea \label{e7323} I_0[a,b,c] = \int
\frac{d^4k}{(2 \pi)^4} \frac{1}{\left[ k^2-m_{c(b)}^2 \right]^a \left[
(p+k)^2-m_{c(b)}^2 \right]^b \left[ (p^\prime+k)^2-m_{c(b)}^2\right]^c}~,
\nnb \\ \nnb \\
I_\mu[a,b,c] = \int \frac{d^4k}{(2 \pi)^4} \frac{k_\mu}{\left[
k^2-m_{c(b)}^2 \right]^a \left[ (p+k)^2-m_{c(b)}^2 \right]^b \left[
(p^\prime+k)^2-m_{c(b)}^2\right]^c}~,
 \eea where $k$ is the momentum of the spectator quark $m_{c(b)}$.

  These integrals are
calculated by shifting  to the Euclidean space--time and using
the Schwinger representation for the Euclidean propagator:
 \bea
\label{e7324} \frac{1}{k^2+m^2} = \frac{1}{\Gamma(\alpha)}
\int_0^\infty d\alpha \, \alpha^{n-1} e^{-\alpha(k^2+m^2)}~.
 \eea
This kind of expression  is very easy for the Borel transformation since \bea
\label{e7325} {\cal B}_{\hat{p}^2} (M^2) e^{-\alpha p^2} = \delta
(1/M^2-\alpha)~. \eea where $M$  is  Borel parameter.

We perform integration over the loop momentum
and over the two parameters which we use in the exponential
representation of propagators \cite{R7321}. As a final operation we  apply double
Borel transformations to $p^2$ and $p^{\prime 2}$. We get the Borel transformed form of the integrals in Eq. (\ref{e7323}) as:

 \bea
\label{e7326} \hat{I_0}(a,b,c)& =&i\frac{(-1)^{a+b+c}}{16
\pi^2\,\Gamma(a) \Gamma(b) \Gamma(c)}
(M_{1}^2)^{2-a-b} (M_2^2)^{2-a-c} \, {\cal U}_0(a+b+c-4,1-c-b)~, \nnb \\ \nnb \\
\hat{I_0}_\mu(a,b,c) &=&\hat{I_1}(a,b,c) p_\mu + \hat{I_2}(a,b,c) p'_\mu~,  \eea where
\begin{eqnarray}
 \hat{I_1}(a,b,c) &=& i \frac{(-1)^{a+b+c+1}}{16
\pi^2\,\Gamma(a) \Gamma(b) \Gamma(c)}
(M_{1}^2)^{2-a-b} (M_2^2)^{3-a-c} \, {\cal U}_0(a+b+c-5,1-c-b)~, \nonumber \\ \nonumber \\
\hat{I_2}(a,b,c) &=& i \frac{(-1)^{a+b+c+1}}{16 \pi^2\,\Gamma(a)
\Gamma(b) \Gamma(c)}
(M_{1)}^2)^{3-a-b} (M_2^2)^{2-a-c} \, {\cal U}_0(a+b+c-5,1-c-b)~,
\end{eqnarray}
  and $M_{1}^2$ and $M_2^2$ are the Borel
parameters. The function ${\cal U}_0(a,b)$ is  as follows:
\bea {\cal U}_0(a,b) = \int_0^\infty dy (y+M_{1}^2+M_2^2)^a y^b
\,exp\left[ -\frac{B_{-1}}{y} - B_0 - B_1 y \right]~, \nnb \eea
where \bea \label{e7328}
B_{-1}& =& \frac{1}{M_{1}^2M_2^2}
\left[m_{c(b)}^2(M_{1}^4+M_2^4) + M_2^2M_{1}^2 (2m_{c(b)}^2-q^2) \right] ~, \nnb \\
B_0 &=& \frac{2m_{c(b)}^2}{M_{1}^2 M_2^2} \left[  M_{1}^2 + M_2^2 \right] ~, \nnb \\
B_{1} &=& \frac{m_{c(b)}^2}{M_{1}^2 M_2^2}~. \eea

where the circumflex of $\hat{I}$ in the equations is used for the results  after the  double Borel transformation.
As a result of the lengthy calculations we obtain the following expressions for the two gluon condensate:
\begin{eqnarray}
\Pi_{nonper}&=&-\frac{2\pi\alpha_s\langle G\rangle^2}{3}16m_{c(b})(-\hat{I_0}(1,2,2)+6\hat{I_0}(1,3,1)+\hat{I_0}(2,1,2)-2(\hat{I_0}(2,2,1)-2\hat{I_1}(1,2,2)\nonumber\\&-&
6\hat{I_ 1}(1,3,1)+2 \hat{I_1}(2,1,2)-6\hat{I_1}(2,2,1)+\hat{I_ 1}(3,1,1)+3m_{c(b)})2(\hat{I_0}(1,1,4)+2(\hat{I_0}(1,4,1)+ \hat{I_0}(4,1,1)\nonumber\\&+&\hat{I_1}(1,1,4)+2( \hat{I_1}(1,4,1)+\hat{I_1}(4,1,1))))-3\hat{I_2}(1,1,3))+6 \hat{I_2}(1,3,1))
\end{eqnarray}

Now, we can compare  $g_{\mu\nu}$ coefficient of  Eq. (\ref{CorrelationFuncPhys2}) and Eq. (\ref{CorrelationFuncQCD}) . Our result related to the  sum rules for the corresponding form factor is as follows:

\begin{eqnarray}\label{CoupCons-YbBB-Bpsoffshel}
F(q^2)&=&\frac{e^{\frac{m_{S}^2}{M^2}}e^{\frac{m_{V}^2}{{M^{\prime}}^2}}}{f_V f_S m_Vm_S}
\left[\frac{1}{4~\pi^2}\int^{s_0}_{4m_{c(b)}^2}
ds\int^{s^{\prime}_0}_{4m_{c(b)}^2} ds^{\prime}
\rho(s,s^{\prime},q^2)\right.
\left. \theta
[1-{(f^{(i)}(s,s^{\prime}))}^2]e^{\frac{-s}{M^2}}e^{\frac{-s^{\prime}}
{{M^{\prime}}^2}}+\Pi_{nonper}\right],
\end{eqnarray}
Note that, finally we have to set $q^2=0$ for the real photon.
\section{Numerical analysis}

In this section we calculate the value of form factors and  the barching ratios. We use,
$m_c= 1.275 \pm 0.025~GeV$, $m_b= 4.65 \pm 0.03~GeV$\cite{PDG2012}, $m_{j/\psi}=3096.916 \pm 0.011 MeV$\cite{PDG2012},
 $m_{\chi_{c0}}= 3414.75 \pm0.31 MeV$\cite{PDG2012},$m_{\chi_{b0}}=  9859.44 \pm 0.42 \pm 0.31~MeV$\cite{PDG2012},$m_{\Upsilon}=  9460.30 \pm 0.26$MeV\cite{PDG2012},
$f_{\chi_{c0}}=(343\pm 112)~MeV$\cite{Veliev:2010gb}, $f_{\chi_{b0}}=(175\pm 55)~MeV$\cite{Veliev:2010gb}, $f_{j/\psi}=(481\pm 36)~MeV$\cite{Veliev:2011kq}, $f_{\Upsilon}=(746\pm 62)~MeV$ \cite{Veliev:2011kq} and 	the full width for $\chi_c$: $\Gamma^{\chi_c}_{tot} = 10.4 \pm 0.6$ MeV\cite{PDG2012}.

To do further numerical analyses we have to know the value or range of the auxiliary parameters of QCD sum rules.   Those are  the continuum thresholds( $s_0$ and $s_0^{\prime}$) and  the Borel mass parameters( $M^2$ and ${M^{\prime}}^2$).  The physical results are required to be either weakly depend on or independent of aforementioned parameters. Therefore, we must consider the  working regions of these auxiliary parameters where the dependence of the form factors  are weak. We also consider the working regions for the Borel mass parameters $M^2$ and
${M^{\prime}}^2$  in a way that both the contributions of the higher states and continuum are sufficiently suppressed and the contributions coming from higher dimensions operators  can be ignored. With the aforementioned conditions, we get, $12~GeV^2\leq M^2\leq 25~GeV^2$ and
 $10~GeV^2\leq M'^2\leq 20~GeV^2$ for \xjga decays and  $15~GeV^2\leq M^2\leq 30~GeV^2$ and
 $15~GeV^2\leq M'^2\leq 30~GeV^2$ for \xUga decays.

The continuum thresholds, $s_0$ and $s_0^{\prime}$ must not be greater than the
energy of the first excited states with the same quantum numbers. In  our numerical calculations  the following regions for the continuum thresholds in $s$ and $s'$ channels are used:
$(m_{S}+0.3)^2\leq s_0\leq(m_{S}+0.7)^2$  and $(m_{V}+0.3)^2\leq s_0^{\prime}\leq (m_{V}+0.7)^2$ for  $s$ and $s^{\prime}$ channels, respectively.
Here, $m_S$ is the mass of either $\chi_{c0}$ or $\chi_{b0}$ meson and  $m_V$ is the mass of either $j/\psi$ or $\Upsilon$ meson.
  Note that, we follow the standard procedure in the QCD sum rules, where the continuum thresholds are suppossed to be independent of Borel mass parameters and $q^2$. However, this assumption is not free of  uncertainties(see for instance \cite{melikhov}).

The detailed analysis of the form factor shows us that dependence of form factor  best fits into the following function:
%
\begin{eqnarray}\label{fitfunction}
F(Q^2)=a e^{-b Q^2}+c
\end{eqnarray}
where we  $a=0.73\pm0.01$, $b=0.2\pm0.01$ and $c=0.012\pm 0.001$ for \xjga and $a=0.4812\pm 0.03$, $b=0.2\pm0.001$ and $c=0.0084\pm0.0003$ for \xUga decays.

Using $Q^2=0$ in Eq. (\ref{fitfunction}), we obtain the $F(0)=0.75\pm0.05~GeV^{-1}$  and the $F(0)=0.47\pm0.03~GeV^{-1}$  for \xjga and  \xUga decays, respectively.

The errors in our numerical calculation are  the results of both  the interval of the working regions for the auxiliary parameters and  the uncertainties of the input parameters.

The matix element for the decay of \xjga and \xUga is as follows:
\begin{eqnarray}
  M &=&e F(q^2=0)\{(p^\prime \cdot q)\epsilon^{\prime} \cdot \epsilon -(q\cdot \epsilon^{\prime})(p^\prime\cdot \epsilon)\}
\end{eqnarray}
where $p^{\prime}$ and $\epsilon^{\prime}$ are the momentum and polarization of final state vector meson i.e., either $j/\psi$ or $\Upsilon$ mesons, $\epsilon$ is the polarization of  real photon and $p$ is the momentum of  initial scalar meson.

Using this matix element, we get:
\begin{equation}\label{decay}
  \Gamma=\frac{|\overrightarrow{p}|}{8 \pi m_S^2}|M|^2=\frac{\alpha}{8}F^2(0)m^3_S(1-\frac{m_V^2}{m_S^2})^3
\end{equation}
where $m_S$ is mass of either $\chi_{c0}$ or $\chi_{b0}$ meson and $m_V$ is either mass of $j/\psi$ or $\Upsilon$ meson.
The decay width for \xjga decays is as:
  \begin{equation}\label{ndecay}
\Gamma (\chi_{c0}\rar j/\psi \gamma)=(11.2\pm 0.3) \times 10^{(-5)}\text{GeV}.
\end{equation}

The branching ratio of \xjga can be evaluated with the Eq. (\ref{ndecay})and the numerical value of $F(0)=0.75\pm0.05$ that is as:
\begin{equation}\label{Br}
{\cal B}_r(\chi_c\rar j/\psi \gamma)=1.07\times 10^{(-2)}\pm 0.14 \times 10^{(-2)}
\end{equation}
this result is in good agreement with the experimental measurement\cite{PDG2012} which is:
\begin{equation}\label{exBr}
{\cal B}_r(\chi_c\rar j/\psi \gamma)=(1.17\times \pm 0.08)\times 10^{(-2)}.
\end{equation}
We get $F(0)=0.47\pm0.03~GeV^{-1}$ for \xUga decays. Using this value we calculate the decay width as follows:
  \begin{equation}\label{exBr}
\Gamma (\chi_{b0}\rar \Upsilon \gamma)=(9.9\pm 0.8) \times 10^{(-5)}\text{GeV}.
\end{equation}
This decay width and the measured branching ratio ${\cal B}_r(\chi_b\rar \Upsilon \gamma)=(1.76 \pm 0.30)\times 10^{(-2)}$\cite{Kornicer:2010cb} allow us to evaluate the total width of $\chi_{b0}$. We estimate that the Full Width $\Gamma_{tot}(\chi_{b0}(1P)) =5.5\pm 0.5 ~MeV$, which is consistent with the experimental results that indicate  the Full Width $\Gamma_{tot}<6~MeV$\cite{Kornicer:2010cb}

\section{Acknowledgement}
The author is grateful to  TM. Aliev and K. Azizi for their useful discussions.


\begin{thebibliography}{99}
 \bibitem{1}N. Brambilla et al., hep-ph/0412158, D. Besson, T. Skwarnicki, Ann. Rev. Nucl. Part. Sci. 43, 333 (1993).
\bibitem{Veliev:2011kq}
 E.~V.~Veliev, K.~Azizi, H.~Sundu, G.~Kaya and A.~Turkan,
  Eur.\ Phys.\ J.\ A {\bf 47}, 110 (2011)
  [arXiv:1103.4330 [hep-ph]].
  \bibitem{Wang:2012kw}
  Z.~-G.~Wang,
  arXiv:1203.6252 [hep-ph].
\bibitem{Rodrigues:2010ed}
  B.~Osorio Rodrigues, M.~E.~Bracco, M.~Nielsen and F.~S.~Navarra,
  Nucl.\ Phys.\ A {\bf 852}, 127 (2011)
  [arXiv:1003.2604 [hep-ph]].


\bibitem{Sundu:2011vz}
  H.~Sundu, J.~Y.~Sungu, S.~Sahin, N.~Yinelek and K.~Azizi,
  Phys.\ Rev.\ D {\bf 83}, 114009 (2011)
  [arXiv:1103.0943 [hep-ph]].



\bibitem{Veliev:2010vd}
  E.~V.~Veliev, K.~Azizi, H.~Sundu and N.~Aksit,
  J.\ Phys.\ G {\bf 39}, 015002 (2012)
  [arXiv:1010.3110 [hep-ph]].
%
\bibitem{pdg1} N. Brambilla et al., CERN-2005-005, (CERN, Geneva,
2005), arXiv:hep-ph/0412158.
\bibitem{pdg2} E. Eichten et al., Rev. Mod. Phys. 80, 1161 (2008)
arXiv:hep-ph/0701208.
\bibitem{pdg3}S. Eidelman, H. Mahlke-Kruger, and C. Patrignani, in
C. Amsler et al. (Particle Data Group), Phys. Lett. B667,
1029 (2008).
\bibitem{pdg4} S. Godfrey and S.L. Olsen, Ann. Rev. Nucl. Part. Sci. 58
51 (2008), arXiv:0801.3867 [hep-ph].
\bibitem{pdg5} T. Barnes and S.L. Olsen, Int. J. Mod. Phys. A24, 305
(2009).
\bibitem{pdg6} G.V. Pakhlova, P.N. Pakhlov, and S.I. Eidelman, Phys.
Usp. 53 219 (2010), [Usp. Fiz. Nauk 180 225 (2010)].
\bibitem{pdg7} N. Brambilla et al., Eur. Phys. J. C71, 1534 (2011)
arXiv:1010.5827 [hep-ph].
\bibitem{Bashiry:2011fr}
  V.~Bashiry,
  Phys.\ Rev.\ D {\bf 84}, 076008 (2011)
  [arXiv:1109.5212 [hep-ph]].

\bibitem{Bashiry:2011pp}
  V.~Bashiry, K.~Azizi and S.~Sultansoy,
  Phys.\ Rev.\ D {\bf 84}, 036006 (2011)
  [arXiv:1104.2879 [hep-ph]].
%
%
%
%
%
\bibitem{R7320} V. A. Fock, Sov. Phys. {\bf 12}, 404 (1937).
\bibitem{R7321} J. Schwinger, Phys. Rev. {\bf 82}, 664 (1951).
\bibitem{R7322} V. Smilga, Sov. J. Nucl. Phys. {\bf 35}, 215 (1982).
\bibitem{Aliev3}   T. M. Aliev, M. Savci, Eur. Phys. J. {\bf C 47} (2006)
413.
\bibitem{R7323} V. V. Kiselev, A. K. Likhoded,  A. I. Onishchenko,
 Nucl. Phys.  {\bf B 569}  (2000) 473.
%
%
\bibitem{PDG2012}J. Beringer et al. (Particle Data Group), Phys. Rev. D86, 010001 (2012)
\bibitem{Veliev:2010gb}
  E.~V.~Veliev, H.~Sundu, K.~Azizi and M.~Bayar,
  Phys.\ Rev.\ D {\bf 82}, 056012 (2010)
  [arXiv:1003.0119 [hep-ph]].

\bibitem{melikhov} W. Lucha, D. Melikhov and S. Simula, Phys. Rev. D
 79, 0960011 (2009).

\bibitem{Kornicer:2010cb}
  M.~Kornicer {\it et al.}  [CLEO Collaboration],
  Phys.\ Rev.\ D {\bf 83}, 054003 (2011)
  [arXiv:1012.0589 [hep-ex]].
\end{thebibliography}
\end{document}